\documentclass[12pt,fleqn]{article}
\usepackage{verbatim,amssymb,amsmath,graphics,amsthm,cite}
\numberwithin{equation}{section}
\newtheorem{proposition}{Proposition}[section]
\newtheorem{theorem}{Theorem}[section]
\newtheorem{corollary}{Corollary}[section]
\newtheorem{definition}{Definition}[section]

\def\<{\langle}
\def\>{\rangle}
\def\+{\phi^+}
\def\-{\psi^-}
\def\l{\left|\!\left|}
\def\r{\right|\!\right|}
\def\b{\begin{equation}}
\def\e{\end{equation}}

\def\la{\label}

\def\A{{\cal A}}
\def\B{{\cal B}}
\def\D{{\cal D}}

\def\G{{\cal G}}
\def\H{{\cal H}}

\def\S{{\cal S}}
\def\T{{\cal T}}
\def\V{{\cal V}}
\def\a{\alpha}
\def\s{{\mathsf{s}}}

\def\Int{{\rm Int}}
\def\ad{{\rm ad}}

\begin{document}
\title{Symmetry Representations in the Rigged Hilbert Space
Formulation of Quantum Mechanics} 
\author{
S.~Wickramasekara\footnote{sujeewa@physics.utexas.edu}\; and A.~Bohm\\ 
Department of Physics, University of Texas at Austin\\
Austin, Texas 78712}
\date{}
\maketitle
\begin{abstract}
We discuss some basic properties of Lie group representations in
rigged Hilbert spaces. In particular, we show that a differentiable
representation in a rigged Hilbert space may be obtained as the
projective limit of a family of continuous representations in a nested
scale of Hilbert spaces. We also construct a couple of examples
illustrative of the key features of group representations in rigged
Hilbert spaces. Finally, we establish a simple criterion for the
integrability of an operator Lie algebra in a rigged Hilbert space. 
\end{abstract}

\section{Introduction}\label{sec1}

In this paper we undertake a study of differentiable representations
of finite dimensional Lie groups in rigged Hilbert spaces (RHS). 
Since symmetry transformations on 
physical systems often constitute such Lie groups, these
representations may prove to be an integral component of the
relatively new rigged Hilbert space formulation of quantum
physics~\cite{roberts,boulder,antione,bohm-jmp,bohm-gadella}. 
The inceptive motivation for introducing RHS in quantum mechanics,
especially in~\cite{roberts, boulder,antione}, was to provide Dirac's
bra and ket formalism, already a well established calculational tool,
with a proper mathematical content. It was later realized
\cite{bohm-jmp,bohm-gadella,physica,harshman} that the mathematical
structure of RHS 
contains a certain suppleness that is well suited for a systematic
study of scattering and decay phenomena. During about the past two
decades, investigations have continued into various aspects of the
quantum theory 
of scattering and decay  
in the framework of RHS. Perhaps the most significant of these
developments is the finding that in a suitably constructed RHS, the
fundamental dynamical equation of Schr\"odinger
$i\hbar\frac{\partial\psi}{\partial t}=H\psi$ can be  integrated to
obtain a Hamiltonian generated {\em semigroup} for the time evolution
of the physical system~\cite{bohm-jmp,bohm-gadella,manolo}. 
This and certain other features of the theory
show that the RHS formulation of quantum physics deviates from the
orthodox Hilbert space theory in significant ways. They are also
indicative of the above mentioned flexibility of the structure of RHS
mathematics.

However, although the semigroup time evolution in RHS has been studied
extensively~\cite{manolo} and often emphasized, a systematic study of
representations of Lie 
groups in RHS has not been carried out in a general setting. Certain
fundamental properties such representations must possess, as well as
their physical content, have been discussed in \cite{antione,
antione2}. Even in these works, some of the most natural questions to
address, such as obtaining an RHS representation of a Lie group from a
given Hilbert space representation and/or from a given Lie algebra
representation, have not been undertaken.

Aside from the rather obvious need as a component of the general RHS
formulation of quantum mechanics, such a study of Lie group
representations in RHS is also motivated by certain recent
applications of the 
formalism to relativistic resonances and unstable particles
\cite{rgv}. These works develop a characterization of relativistic
resonances and unstable particles by way of certain representations of
a particular subsemigroup of the Poincar\'e group.  The relevant
subsemigroup, named Poincar\'e semigroup \cite{rgv}, is in fact the
semidirect product of the homogeneous 
Lorentz group with the {\em semigroup} of space-time translations into
the forward light cone. The RHS representations of this subsemigroup
can be characterized by a spin value $j$ and a complex square mass
value $\s_R$, and consequently they can be attributed a physical
interpretation as representing resonances along the lines of Wigner's
classic theory of the unitary representations of Poincar\'e group
for stable particles. These RHS representations of the Poincar\'e
semigroup have subtleties which are not present in either the unitary
representations in Hilbert spaces or the well understood
RHS theory for the non-relativistic case \cite{manolo} where only a
one parameter 
semigroup is needed to describe the evolution of the physical
system. Many of the technical and theoretical issues appertaining to
the RHS representations of the multi-parameter Poincar\'e
semigroup~\cite{rgv} are subsumed under the subject of this paper.
In the remainder of this introductory section we shall briefly state
the questions that we attempt to formulate and answer in this paper;
in Sections \ref{sec2} and \ref{sec3} we present our results.

\begin{definition}\label{def1.1} 
A rigged Hilbert space consists of a triad of vector spaces
\begin{equation}
\Phi\subset\H\subset\Phi^\times\label{1.1}
\end{equation}
where: 
\begin{enumerate}
\item
$\cal H$ is a Hilbert space
\item
$\Phi$ is a dense subspace of $\cal H$ 
and it is  endowed with a complete
locally convex topology $\tau_\Phi$ that is stronger than the 
$\H$-topology 
\item
$\Phi^\times$ is the
space of continuous antilinear functionals on $\Phi$. It is complete 
in its weak* topology 
$\tau^\times$ and it contains $\H$ as a  dense subspace. 
\end{enumerate}
\end{definition}

It is preceptive that 
the topology of the space $\Phi$ 
be constructed so as to yield an algebra $\A$ of quantum
physical observables --defined at the outset as an algebra of
endomorphisms on a dense subspace $\D$ of $\cal H$-- continuous as mappings
on $\Phi$.  For an operator $A$ of this algebra 
(that is also self adjoint, normal or
unitary as an operator in $\cal H$), the Nuclear Spectral Theorem of
Gel'fand affirms the existence of generalized eigenvectors (i.e.,
eigenvectors of the dual operator $A^\times$ in
$\Phi^\times$) with the corresponding eigenvalues ranging over the
continuous (Hilbert space) spectrum of
$A$\footnote{In Gel'fand's original proof of the theorem, the locally
convex space $\Phi$ of Definition~\ref{def1.1} was required to be
nuclear. Therefore, rigged Hilbert spaces are customarily defined in
quantum theory with the requirement that $\Phi$ be nuclear. However,
since this condition can be relaxed~\cite{antione} and since the
nuclearity of $\Phi$ is not needed for the purposes of this paper (and
thus our results have a slightly broader generality), we
choose to define RHS's as in Definition~\ref{def1.1}, without demanding
that $\Phi$ be nuclear. See also \cite{pietsch}.\label{fn}}. 

Thus, with the aid of
RHS, the continuous and point spectra of observables can be treated on an
equal footing. Further, the above set of eigenvectors constitute a
basis for the space $\Phi$. This is in fact the mathematical content
of Dirac's bra-and-ket formulation of quantum mechanics.

Very often in practice, the above mentioned algebra of observables
$\A$ (to be made continuous on $\Phi$) arises as the associative
algebra of an operator Lie algebra in $\H$. Further, this Lie algebra
may be the differential $d\T$ (with respect to the norm topology of
$\H$) of a continuous (often unitary) representation $\T$ in $\H$ of a
Lie group $G$. As stated above, the complete locally convex space $\Phi$ for
an RHS may be constructed from an invariant dense domain $\D$ for the
associative algebra of $d\T$ so that every element of this algebra
becomes continuous as a mapping on $\Phi$.

We prove (Proposition \ref{prop2.1}) that the natural question whether
the Hilbert space representation $\T$ (say, when restricted to $\Phi$)
yields a representation of the group $G$ in $\Phi$ is answerable in
the affirmative, provided the invariant domain for the operator Lie
algebra $d\T$ is chosen so that it remains invariant also under the
group representation $\T$. Observe that this is a natural and minimal
requirement for a homomorphism to be defined on $G$ by the composition
of the operators $\T|_\Phi$ which denote the restriction of $\T$ to
$\Phi$. Moreover, it will be seen that the $\tau_\Phi$-generators of
the representation $\T|_\Phi$ coincide with the $\tau_\H$-generators
of $\T$ on the space $\Phi$. 

In contrast, it may also be possible to construct the space $\Phi$
from a dense domain $\D$ which remains invariant under the
differential $d\T$ but not under the group representation $\T$. This
leads to the interesting possibility that certain symmetries 
present in the Hilbert space description of a quantum mechanical
system need not be present in its  
RHS description. It is this feature that has
been exploited in the above mentioned RHS 
study of certain quantum mechanical processes such as resonance
scattering and decay, and in particular,    
the apparent asymmetric, semigroup time evolution associated to these
processes. However, we shall not be concerned with these aspects of
the RHS quantum theory in this paper.

Section~\ref{sec3} of this paper deals with the 
complementary question whether  every (differentiable)
Lie group representation in the space $\Phi$ of an RHS is necessarily
obtained from a (continuous) representation of the group in the
central Hilbert space $\H$. The starting point in this case 
is a representation
$T$ of a certain Lie algebra $\G$ in a Hilbert space $\H$. Unlike in
Section~\ref{sec2}, Proposition~\ref{prop2.1}, 
we will no longer assume that $T$ is the
differential $d\T$ of a continuous group representation $\T$ in
$\H$. Instead, we will establish a simple criterion of determining if
the given Lie algebra representation $T$ is the differential of a
certain Lie group representation in $\Phi$.

\section{Induction from Hilbert Space Representations}\label{sec2}
\begin{definition}\label{def2.1}
A continuous representation of a Lie group $G$ on a topological vector
space $\Psi$ is a continuous mapping 
$\T:\ G\times\Psi\rightarrow\Psi$ such that
\begin{enumerate}
\item
for every $g\in G$, $\T(g)$ is a linear operator in $\Psi$
\item
for every $\psi\in\Psi$ and $g_1,\,g_2\in G$,
$\T(g_1g_2)\psi=\T(g_1)\T(g_2)\psi$
\end{enumerate}
\end{definition}
\begin{definition}\label{def2.2}
A differentiable representation of a Lie group $G$ on a complete 
topological
vector space $\Psi$ is a mapping $\T:\ G\times\Psi\rightarrow\Psi$
which fulfills all the requirements of Definition \ref{def2.1} and has
the additional property that for any one parameter subgroup $\{g(t)\}$ of $G$,
$\lim_{t\rightarrow0}\frac{\T(g(t))\phi-\phi}{t}$ exists for all
$\phi\in\Psi$ (and, {\rm a fortiori}, defines a continuous linear
operator on $\Psi$).  
\end{definition}
\begin{definition}\label{def2.3}
A continuous one parameter group of operators $\T(t)$
in a locally convex topological vector space $\Psi$ is said to be
equicontinuous if for every continuous seminorm $p$ on 
$\Psi$, there exists another, $q$, such
that
\begin{equation}
p(\T(t)\phi)\leq q(\phi)\label{2.3}
\end{equation}
holds for all $\phi\in\Psi$ and all $t\in\mathbb{R}$.\\
The one parameter group is said to be locally equicontinuous if
\eqref{2.3} holds for all $t$ in every compact subset of $\mathbb{R}$.
\end{definition}

Let $G$ be a Lie group of dimension $d<\infty$, and $\G$ be its Lie
algebra. Let $\T$ be a continuous representation of $G$ 
in a Hilbert space $\H$, and let $T$ be the differential of $\T$
evaluated at the identity $e$ of $G$, $d\T|_e=T$. It is well known
that $T$ furnishes a representation of $\G$ by (not necessarily continuous)
linear operators in $\H$.
\begin{proposition}\label{prop2.1}
Let $G,\ \G,\ \T$ and $T$ be as above. Let $\D$ be a dense subspace of
$\H$ which remains invariant under both $\T$ and $T$.   
Then there exists a rigged Hilbert space
$\Phi\subset\H\subset\Phi^\times$ such that  
the restrictions ${\T}\vert_{\Phi}$ 
yield a continuous 
representation of $G$ in $\Phi$.\\
Furthermore, if $\D$ can be chosen so that it is complete under the
projective topology $\tau_\Phi$ (\eqref{2.5} below), the
representation $\T|_\Phi$ of $G$ is differentiable in $\Phi$.
By duality, there also exists a
differentiable representation of $G$ in $\Phi^\times$.\\
\end{proposition}
\vskip .2cm
\noindent{\footnotesize PROOF:}\\
Let $\{x_i\}_{i=1}^d$ be a basis for $\G$ and let $X_i$ be  the
restriction of the differential $T(x_i)$ to the invariant domain
$\D$. 
\vskip .5cm
\noindent{\bf Construction of RHS}\\
Define a family of scalar
products on $\D$ by setting
\begin{equation}
(\phi,\psi)_{n+1}=\sum_{i=1}^d(X_i\phi,X_i\psi)_n+(\phi,\psi)_n,\quad
n=0,1,2,\cdots,\ \phi,\psi\in\D\label{2.4}
\end{equation}
where $(\phi,\psi)_0\equiv(\phi,\psi)$, the scalar product which $\D$
inherits from $\H$. Linearity of the $X_i$ then ensures that
$(\phi,\psi)_n$ is in fact a scalar product on $\D$ for every $n$. 

With \eqref{2.4}, we have on $\D$ the family of norms,
\begin{equation}
\l\phi\r_{n+1}^2=\sum_{i=1}^d\l X_i\phi\r_n^2+\l\phi\r_n^2\la{2.5}
\end{equation}

From \eqref{2.5}, it is clear 
\begin{equation}
\l\phi\r_n\leq\l\phi\r_{n+1}\quad\text{and}\quad\l
X_i\phi\r_n\leq\l\phi\r_{n+1}\label{2.6}
\end{equation}

Since the norms \eqref{2.5} are derived from the scalar products
\eqref{2.4}, the dense subspace $\D$ can be completed with respect to
each norm $\l.\r_n$ to obtain a Hilbert space $\H_n$. The relations
\eqref{2.6} then imply that the $\H_n$ form a nested scale 
\begin{equation}
\H\supset\H_1\supset\H_2\supset\cdots\label{2.7}
\end{equation}
and that the operators $X_i$, and therewith the algebra $\A$ spanned
by them, extend to elements of $\B(\H_{n+1},\H_n)$, the space of
bounded linear operators from $\H_{n+1}$ into $\H_n$. 

Now, let $\Phi$ be defined by
\begin{equation}
\Phi=\bigcap_n\H_n\label{2.8}
\end{equation}
It is clear that $\Phi$ is a Fr\'echet space\footnote{The topology of
$\Phi$ induced by the countable family of norms \eqref{2.5} is
equivalent to the topology induced by the powers of 
generalized Laplacian $(\sum_{i=1}^dX_i^2+I)^n$, as considered in
\cite{bohm-semisimple}.} which contains $\D$. It is also easy to see
that the topology of $\Phi$ is independent of the basis chosen. $\Phi$
is dense in $\H$, and thus we have the triplet
\begin{equation}
\Phi\subset\H\subset\Phi^\times\label{2.9}
\end{equation}
where $\Phi^\times$, the anti-dual of $\Phi$, can be obtained as
\begin{equation}
\Phi^\times=\bigcup_{n}\H_n\label{2.9b} 
\end{equation}
\noindent{\bf Remark}\label{remark}\\
It is not known to us if the space $\Phi$ is nuclear 
when it is constructed in the manner above, i.e., under the
projective topology from the differential of a 
continuous representation of a finite dimensional but otherwise
arbitrary Lie group in a Hilbert space. However,
it is known that nuclearity holds for $\tau_\Phi$ for the unitary
representations of the following classes of Lie groups:
semi-simple groups~\cite{bohm-semisimple}; nilpotent
groups~\cite{nagel}; semi-direct products of Abelian groups with compact
groups~\cite{nagel}; and the Poincar\'e group. 
Thus for a large class of Lie groups, our Proposition \ref{prop2.1}
can be restated  for a triad $\Phi\subset\H\subset\Phi^\times$, where
$\Phi$ is a nuclear space. 
\vskip .5cm
\noindent{\bf Restriction of $\boldsymbol{\T}$ to $\boldsymbol{\Phi}$} 

From the $\H$-continuity of $\T(g)$, we have, for all $\psi\in\H$,
\begin{equation}
\l\T(g)\psi\r\leq\omega(g)\l\psi\r\label{2.10}
\end{equation}
where $\omega(g)$ is a positive constant which may depend on the group
element $g$. An important property of the representation $\T$ is that
it is locally equicontinuous, a consequence of the local
equicontinuity of continuous, one parameter groups in barrelled
spaces~\cite{komura}. That is, the positive valued function $\omega$
on $G$ is locally bounded.

Proposition \ref{prop2.1} follows from \eqref{2.10} and the
following operator valued formulation of the well known Lie algebra
inner automorphism ${\rm Ad}(e^{ty})$ of $\G$, 
defined by $z\rightarrow
e^{ty}ze^{-ty},\ y,z\in\G$ (in any realization). Thus, for $g=e^y$,
\begin{equation}
gzg^{-1}=e^{({\rm ad}y)}z\equiv f_{zi}(g^{-1})x_i\label{2.11}
\end{equation}
where the functions $f_{zi}$ are locally analytic on $G$. The
corresponding automorphism on $G$ is $ge^{tz}g^{-1}=e^{(t{\rm
exp}({\rm ad}y)z)}$, where $g=e^y$ and $t$, a real parameter. Then,
for $\phi\in\D$, 
\begin{equation}
\frac{d}{dt}\T(g)\T(e^{tz})\T(g^{-1})\phi=\frac{d}{dt}\T(e^{t({\rm
exp}({\rm ad}y)z)})\phi\label{2.12}
\end{equation}
Now, since
\begin{eqnarray}
\lefteqn
{\lim_{t\rightarrow0}
\left|\!\left|\frac{{\T}(g){\T}(e^{tz}){\T}(g^{-1})\phi-\phi}{t}-{\T}(g)T(z)
{\T}(g^{-1})\phi\right|\!\right|}\quad\qquad\qquad
\nonumber\\
&&\leq
\omega(g)\lim_{t\rightarrow0}
\left|\!\left|\left(\frac{{\T}(e^{tz})-I}{t}-T(z)\right){\T}(g^{-1})\phi
\right|\!\right|\qquad\la{2.13}
\end{eqnarray}
and since $\D$ is invariant under $\T$, we see that the left hand
side of \eqref{2.12}, evaluated at $t=0$, is
$\T(g)T(z)\T(g^{-1})$. Thus,
\begin{equation}
\T(g)T(z)\T(g^{-1})\phi=T((e^{{\rm ad}y})z)\phi\label{2.14}
\end{equation}
for $\phi\in\D$. But, by \eqref{2.11}, for the basis elements $X_i$ 
we then have 
\begin{equation}
\T(g)X_i\T(g^{-1})\phi=\sum_{j=1}^df_{ij}(g^{-1})X_j\phi\label{2.15}
\end{equation}

The real valued functions $f_{ij}$ are continuous and locally
analytic, and provide a (not necessarily faithful) matrix
representation of $G$. For the one parameter subgroup $\{e^{tx_k}\}$,
it is easy to see that the $f_{ij}$ can be expanded as 
\begin{equation}
f_{ij}(e^{-tx_k})=\delta_{ij}+tc_{ijk}+\cdots\label{*}
\end{equation}
where $c_{ijk}$ are the structure constants of $\G$. Furthermore, the
$f_{ij}$ and $c_{ijk}$ fulfill the identities
\begin{equation}
\sum_kc_{ijk}f_{kl}(g^{-1})=\sum_{m,n}c_{mnl}f_{im}(g^{-1})f_{jn}(g^{-1})
\label{**}
\end{equation}
 
The relations \eqref{2.10} and \eqref{2.15} show that for any
$\phi\in\D$,
\begin{equation}
\l\T(g)\phi\r_n\leq\omega(g)\left(1+\sum_{i,j=1}^d|f_{ij}(g)|\right)^n
\l\phi\r_n\label{2.16} 
\end{equation}
The proof of \eqref{2.16} is by induction. For $n=0$, \eqref{2.16} is
just \eqref{2.10}, the assumed continuity of $\T$ in $\H$. If
\eqref{2.16} holds for some $n$, then,
{\footnotesize 
\begin{eqnarray}   
\left|\!\left|{\T}(g)\phi\right|\!\right|_{n+1}^2
&=&\sum_{i=1}^d\left|\!\left|X_i{\T}(g)\phi\right|\!\right|_n^2
+\left|\!\left|{\T}(g)\phi\right|\!\right|_n^2  
\nonumber \\[0.0703cm]
&=&\sum_{i=1}^d\left|\!\left|{\T}(g){\T}(g^{-1}) X_i{\T}(g)\phi
\right|\!\right|_n^2 +\left|\!\left|{\T}(g)\phi\right|\!\right|_n^2
\nonumber \\[0.0703cm]
&\leq&\omega(g)^2\left(1+\sum_{i,j=1}^d|f_{ij}(g)|\right)^{2n}
\left(\sum_{k=1}^d\left|\!\left|{\T}(g^{-1})X_k{\T}(g)
\phi\right|\!\right|_n^2 +\left|\!\left|\phi\right|\!\right|_n^2\right)
\nonumber \\[0.0703cm]
&\leq&\omega(g)^2\left(1+\sum_{i,j=1}^d|f_{ij}(g)|\right)^{2n}
\left(1+\sum_{k,l=1}^d|f_{kl}(g)|\right)^2\left|\!\left|\phi
\right|\!\right|_{n+1}^2
\nonumber \\[0.0703cm]
&\leq&\omega(g)^2\left(1+\sum_{i,j=1}^d|f_{ij}(g)|\right)^{2n+2}
\left|\!\left|\phi\right|\!\right|_{n+1}^2,\label{2.17}
\end{eqnarray}}
where the inequalities \eqref{2.6} are used in the last step. Thus,
we have \eqref{2.16}.

The relation \eqref{2.16} gives the continuity of the operators $\T(g),\ 
g\in G,$ (when restricted to the dense domain $\D$) with respect to the
Fr\'echet topology given by \eqref{2.4} or \eqref{2.5}. It is also
fairly straightforward to establish the continuity of the mapping
$G\rightarrow\T(G)$ in this topology on $\D$. To that end, for
$\phi\in\D$,
\begin{equation}
\l\T(g)\phi-\phi\r_{n+1}^2=\sum_{i=1}^d\l X_i\T(g)\phi-X_i\phi\r_n^2
+\l\T(g)\phi-\phi\r_n^2\la{2.18}
\end{equation}
Then, since{\footnotesize
\begin{eqnarray}
\l X_i\T(g)\phi-X_i\phi\r_n
&=&\l\T(g)\sum_{i,j=1}^df_{ij}(g)X_j\phi-X_i\phi\r_n   
\nonumber \\[0.0703cm]
&\leq&\omega(g)\left(1+\vert\sum_{i,j=1}^df_{ij}(g)\vert\right)^n
\l\sum_{i,j=1}^df_{ij}(g)X_j\phi-X_i\phi\r_n
\nonumber \\[0.0703cm]
&&+
\l\T(g)X_i\phi-X_i\phi\r_n\la{P21}
\end{eqnarray}}
and since from \eqref{*}, $\lim_{g\rightarrow
e}f_{ij}(g^{-1})=\delta_{ij}$, the continuity
$\lim_{t\rightarrow0}\l\T(e^{tx})\phi-\phi\r_n=0$ implies
\begin{equation}
\lim_{t\rightarrow0}\l\T(e^{tx})\phi-\phi\r_{n+1}=0\label{2.19}
\end{equation}
Since $\D$ is dense in each Hilbert space $\H_n$ of the nested scale
\eqref{2.7}, linearity of the operators $\T(g)$ permits the
inequalities \eqref{2.16} and \eqref{2.19} to be extended to the whole
of $\H_n$. That is, the representation $\T|_{\D}$ extends from $\D$ to
a continuous representation of $G$ in each of the Hilbert spaces
$\H_n$ of \eqref{2.7}. 

Since $\Phi=\bigcap_{n=0}^\infty\H_n$, the relations \eqref{2.16} and
\eqref{2.19} can be extended to the space $\Phi$. 
Therewith we
conclude that the restrictions $\T(g)|_\Phi$ to the space $\Phi$
yields a continuous (with respect to the $\Phi$-topology \eqref{2.5})
representation of $G$ on $\Phi$. 

It remains to prove that this representation on $\Phi$ is
differentiable, i.e., for any $\phi\in\Phi$ and $x\in\G$,
$\lim_{t\rightarrow0}\frac{\T(e^{tx})-I}{t}\phi$ exists. We shall
shortly see that the equality
\begin{equation}
\lim_{t\rightarrow0}\l\frac{\T(e^{tx})-I}{t}\phi-T(x)\phi\r_n=0\label{2.20}
\end{equation}
can be easily obtained by induction so long as $\phi$ is restricted to
the dense domain $\D$.  However, since the mapping $G\rightarrow\T(G)$
is not linear, we cannot necessarily extend \eqref{2.20} to the whole
of $\Phi$. 

At this point we remark that a result of Roberts, Proposition 13 in 
\cite{roberts}, leads to the conclusion 
that the invariant domain $\D$ is complete
under the projective topology when $\D$ is taken to be the maximal
invariant domain for the operator Lie algebra $T(\G)$. This domain is
also invariant under the operator group $\T(G)$. Thus, for such $\D$, 
\eqref{2.20} holds for all $\phi\in\Phi$, and we have a differentiable
representation of $G$ on $\Phi$.

To prove \eqref{2.20}, notice first that for $n=0$ the equation 
just expresses that differentiability of $\phi$ in $\H$-topology, and
thus the equation is true for all $\phi\in\D$ by the definition of
$\D$. Next, if \eqref{2.20} is true for some $n$, then

{{\footnotesize 
\begin{eqnarray}
\lefteqn{\lim_{t\rightarrow0}\l\frac{\T(e^{tx_i})-I}{t}\phi-T(x_i)\phi\r^2_{n+1}}
\nonumber\\[0.0703cm]
&&=
\lim_{t\rightarrow0}\left(\sum_{j=1}^d\l
X_j\left(\frac{\T(e^{tx_i})-I}{t}\phi-X_i\phi\right)\r_n^2
+\l\frac{\T(e^{tx_i})-I}{t}\phi-X_i\phi\r_n^2\right)
\label{2.21}
\end{eqnarray}}}
Since \eqref{2.20} is assumed to be true for $n$, the last term
vanishes. Also,
\begin{eqnarray}
\lefteqn{
\l X_j\left(\frac{\T(e^{tx_i})-I}{t}-X_i\right)\phi\r_n}
\qquad\qquad\qquad\qquad\nonumber\\[0.0703cm]
&&=\l\frac{\sum_kf_{jk}(e^{tx_i})
\T(e^{tx_i})X_k\phi-X_j\phi}{t}-X_jX_i\phi\r_n\nonumber\\[0.0703cm]
&&\leq\l\frac{f_{jj}(e^{tx_i})
\T(e^{tx_i})X_j\phi-X_j\phi}{t}-X_iX_j\phi\r_n\nonumber\\[0.0703cm]
&&\quad+\l\frac{\sum_{k\not=j}\left(f_{jk}(e^{tx_i})
\T(e^{tx_i})X_k\phi-tc_{jik}X_k\phi\right)}{t}\r_n\nonumber\\
\label{2.22}
\end{eqnarray}
The invariance of $\D$ under $X_k$ and the expansion \eqref{*} of the
$f_{ij}$ show that the right hand side of \eqref{2.22} vanishes when
$t\rightarrow0$. That is, the right hand side of \eqref{2.21} tends to
zero. This proves \eqref{2.20} for every basis element $x_i$ of
$\G$. The general case easily follows.\\
The existence of a differentiable representation of $G$ in
$\Phi^\times$ easily follows from the treatment in
Section~\ref{sec2.4}.\\
This concludes the proof Proposition \ref{prop2.1}.\hfill$\Box$

\vskip .5cm

Proposition \ref{prop2.1} thus shows that, starting from a continuous
representation of a finite dimensional Lie group in a Hilbert space
$\H$, a 
rigged Hilbert space $\Phi\subset\H\subset\Phi^\times$ can be
constructed so that there exists a differentiable representation of
the group in $\Phi$. The construction begins with identifying the
maximal invariant domain for the operator Lie algebra in $\H$. 
In view of the remark on page \pageref{remark},
for unitary representations of a large class of Lie groups we can
construct the triad $\Phi\subset\H\subset\Phi^\times$ subject to the
more restrictive condition that $\Phi$ be a nuclear space.

In the remainder of this Section we shall investigate some secondary
aspects of such representations in $\Phi$ and present a couple of
simple examples illustrating of these features.

\subsection{One Parameter Subgroups in $\boldsymbol{\Phi}$}\label{sec2.1}
Proposition \ref{prop2.1} asserts that the differentiable
representation $\T_\Phi$ of a finite dimensional Lie group $G$,
obtained from its  continuous Hilbert space representation $\T$,  
is precisely the projective limit of a family of continuous representations
in the nested scale of Hilbert spaces $\H_n$ in \eqref{2.7}. That is,
the representation $\T_\Phi$ in $\Phi$ extends to a continuous 
representation $\T_n\ (\T_0=\T)$ of $G$ in $\H_n$ for
$n=0,1,2,\cdots$. The generators $X_{i,n}$ of the one parameter
subgroups $\T_n(e^{tx_i})$ are the extensions to $\H_n$ by closure, with
respect to the norm topology $\l.\r_n$, of the operators $X_i$
in $\Phi$, and they furnish a representation of the Lie
algebra $\G$ in some algebra $\A(\D_n)$ of endomorphisms on a dense
subspace $\D_n$ of $\H_n$. In fact, the invariant subspace
$\D$ from which the Fr\'echet space $\Phi$ was constructed can
function as $\D_n$ in each $\H_n$.

This observation motivates us to consider the problem of integrating
the Lie algebra representation $T(\G)$ in $\Phi$ to the differentiable
group representation $\T_\Phi$ as, somewhat loosely put, the
projective limit of the integrability problem in the Hilbert spaces
$\H_n$. We shall take up this integrability of an operator Lie algebra
in $\Phi$ as a substantive problem below in Section \ref{sec3}. 
Here we will limit ourselves to the integrability conditions on
a single element of $T(\G)$ into a differentiable one parameter group
in $\Phi$. More precisely, the integrability of an element $X$ of the
continuous Lie algebra representation $T$ in $\Phi$ to a
differentiable one parameter group can be treated as a repeated
application of the classical Hille-Yosida~\cite{hille,yosida} theory
of one parameter $C_0$-groups in Banach spaces. 

Consider again the case studied in Proposition~\ref{prop2.1}. Let us
denote a typical one parameter 
subgroup of this differentiable representation by
$\T_{\Phi}(t,X)$, where $X$ is the generator of $\T_{\Phi}(t,X)$. As
seen from \eqref{2.16}, the differentiable subgroup $\T_\Phi(t,X)$
extends to a $C_0$-group in each of the Hilbert spaces $\H_n$. In
$\H_n$, this subgroup is generated by ${\bar{X}}_n$, the extension to
$\H_n$, by closure, of the operator $X$ in $\Phi$.  If we denote this  
$C_0$-group in $\H_n$ by $\T(t,{\bar{X}}_n)$, then $\T_\Phi(t,X)$
in $\Phi$ is the projective limit of the $C_0$-groups
$\T(t,{\bar{X}}_n)$ in $\H_n$. 

Suppose $\T(t,{\bar{X}}_n)$ is of type $\omega_n$~\cite{hille,yosida},
i.e.,  
\begin{equation}
\omega_n=\inf_{t\not=0}\frac{1}{|t|}\ln\l\T(t,{\bar{X}}_n)\r_n
=\pm\lim_{t\rightarrow\pm\infty}\frac{1}{|t|}
\ln\l\T(t,{\bar{X}}_n)\r_n\label{2.1.6}
\end{equation}

The classical Hille-Yosida theory affirms the following relationship
between the resolvent $R(\lambda,{\bar{X}}_n)$ of ${\bar{X}}_n$ and
the $C_0$-group $\T(t,{\bar{X}}_n)$ generated by ${\bar{X}}_n$:
\begin{eqnarray}
R(\lambda,{\bar{X}}_n)\phi
&=&\int_0^\infty dt e^{-\lambda t}\T(t,{\bar{X}}_n)\phi,\qquad 
\lambda>\omega_n\nonumber\\
R(\lambda,{\bar{X}}_n)\phi
&=&-\int_{-\infty}^0 dt e^{-\lambda t}\T(t,{\bar{X}}_n)\phi,\qquad 
\lambda<-\omega_n
\label{2.1.7}
\end{eqnarray}
\begin{eqnarray}  
\T(t,{\bar{X}}_n)\phi&=&\lim_{\lambda\rightarrow\infty}e^{-\lambda
t}\sum_{j=0}^\infty\frac{(\lambda t)^j}{j!}\left(\lambda
R(\lambda,{\bar{X}}_n)\right)^j\phi,\quad t>0\nonumber\\
\T(t,{\bar{X}}_n)\phi&=&\lim_{\lambda\rightarrow-\infty}e^{-\lambda
t}\sum_{j=0}^\infty\frac{(\lambda t)^j}{j!}\left(\lambda
R(\lambda,{\bar{X}}_n)\right)^j\phi,\quad t<0
\label{2.1.8}
\end{eqnarray}
where all limits are with respect to the $\H_n$-topology. Further, for
some positive $M_n$ and $\beta_n>\omega_n$, we have  
\begin{equation}
\l \left(R(\lambda,{\bar{X}}_n)\right)^p\r_n\leq
M_n(|\lambda|-\beta_n)^{-p}\label{2.1.9}
\end{equation}
for all $\lambda>\beta_n$ and $p=1,2,3,\cdots$.
In fact, the relation \eqref{2.1.9} is a necessary and sufficient
requirement for the closed operator ${\bar{X}}_n$ to generate the
$C_0$-group $\T(t,{\bar{X}}_n)$ in the Hilbert space $\H_n$. 

Since the differentiable subgroup $\T_\Phi(t,X)$ in $\Phi$ is the
projective limit of the continuous groups $\T_\Phi(t,{\bar{X}}_n)$,
we see that the continuous operator $X$ generates a one parameter
group in $\Phi$ when its closure ${\bar X}_n$ fulfills the
relation \eqref{2.1.9} for all $n=0,1,2,\cdots$. That is, for the
kind of differentiable subgroup considered here, the problem of
reconstructing the $\T_\Phi(t,X)$ in terms of (the resolvent of) $X$ in
$\Phi$ can be reduced to the corresponding problem in each of the
$\H_n$ in the nested scale of Hilbert spaces \eqref{2.7}.  

It is interesting at this point to ask if the subgroup $\T_\Phi(t,X)$
can be recovered from its generator $X$ in $\Phi$ 
without appealing to the Banach space theory applied to the Hilbert
spaces $\H_n$. The theory of $C_0$-groups in more general locally
convex spaces has also been developed~\cite{yosida}, and the form of
this general theory is similar to the Banach space theory when the
group is equicontinuous in the parameter. For such a $C_0$-group in a
locally convex space, the resolvent operator of the generator can be
obtained much the same way as in \eqref{2.1.7} as the Laplace
transform of the group. The group, in turn, can be recovered
from the resolvent by way of a limiting process similar to
\eqref{2.1.8}. Of course the integrals and limit processes are now to
be defined with respect to the locally convex topology of the vector
space.

Nevertheless, as evident from the example below, 
such global equicontinuity of may prove to be
too strong a restriction for $C_0$-groups in rigged Hilbert 
spaces. In such situations, the resolvent operator $R(\lambda,X)$ may
fail to exist anywhere in the complex plane, and further, even when it
does exist for all large $|\lambda|$, the group may not be able to
be constructed from 
it as in \eqref{2.1.8}\footnote{As remarked earlier, 
one parameter $C_0$-groups in
$\Phi$ are necessarily 
locally equicontinuous, and these groups have been studied in the
literature~\cite{komura}. However, we shall not make use of the results 
of~\cite{komura} as the structure of $\Phi$, defined by~\eqref{2.8},
makes the case considerably simpler for one parameter 
groups in rigged Hilbert spaces.}. 

One obvious condition under which the resolvent operator $R(\lambda,
X)$ can acquire an integral resolution of the kind \eqref{2.1.7} in
$\Phi$ is 
\begin{equation}
\omega\equiv\sup_n\omega_n<\infty\label{2.1.10}
\end{equation}
where the $\omega_n$ are defined as in \eqref{2.1.6} and
$|\lambda>|\omega$. However, even when the resolvent $R(\lambda,X)$ of
$X$ is everywhere defined in the
complex plane, it is not necessary that the subgroup $\T_\Phi(t,X)$
can be
recovered in terms of $R(\lambda,X)$ by the limit process
\eqref{2.1.8} (in the $\Phi$-topology). One instance when this is
possible is 
\begin{equation}
M_n\leq1\qquad{\rm
and}\qquad\beta\equiv\sup_n\beta_n<\infty\label{2.1.11}
\end{equation}
where $M_n$ and $\beta_n$ are defined as in \eqref{2.1.9}. This
condition assures that the
Hille-Yosida theory for the $C_0$-groups in locally convex spaces
\cite{yosida} can be applied. In other words, if the relations
\eqref{2.1.10} and 
\eqref{2.1.11} hold, the subgroup $\T_\Phi(t,X)$ can be recovered from
the resolvent of its generator by way of \eqref{2.1.8}, defined now in
$\Phi$ as a $\tau_\Phi$-limit process. 

\subsection{Example}\label{eg1}
Define a multiplication in ${\mathbb{R}}^3$ by
\begin{equation}
(\xi_1,\xi_2,\xi_3)(\zeta_1,\zeta_2,\zeta_3)
=(\xi_1+\zeta_1,\xi_2+\zeta_2,\xi_3+\zeta_3+\xi_1\zeta_2)\label{e1.1}
\end{equation}   
Under this multiplication ${\mathbb{R}}^3$ becomes a group, $G$, which
has the set $\{(0,0,\xi_3)\}$ as its center. The Lie algebra $\G$ of
$G$ is spanned by the elements
\begin{equation}
\chi_1=(1,0,0)\quad\chi_2=(0,1,0)\quad\chi_3=(0,0,1)\label{e1.2}
\end{equation}
which fulfill the commutation relations
\begin{equation}
[\chi_1,\chi_2]=\chi_3,\qquad[\chi_1,\chi_3]=[\chi_2,\chi_3]=0\label{e1.3}
\end{equation}
These commutation relations can be realized in ${\mathbb{R}}^3$ by
the multiplication rule defined, for any two elements 
$\chi=(\alpha,\beta,\gamma)$ and $\chi'=(a,b,c)$ of
$\G$, as
\begin{equation}
(\alpha,\beta,\gamma)(a,b,c)=(0,0,\alpha b)\label{e1.3a}
\end{equation}
Thus, the basis elements \eqref{e1.2} fulfill the relations
\begin{equation}
\chi_i\chi_j=\delta_{1i}\delta_{2j}\chi_3\label{e1.4}
\end{equation}
Notice that under the product rule \eqref{e1.3a}, the Lie algebra $\G$
becomes an associative algebra. This associative algebra can be made
into an operator algebra on ${\mathbb{R}}^3$ by way of the definition, 
for $\chi=(\alpha,\beta,\gamma)\in\G$ and
$v=(x,y,z)\in{\mathbb{R}}^3$, 
\begin{equation}
\chi v=(\alpha y+\gamma z,\ \beta z,\ 0)\label{e1.4b} 
\end{equation}

The group $G$ can be constructed by the exponentiation of $\G$:
\begin{equation}
(\xi_1,\xi_2,\xi_3)=e+\xi_1\chi_1+\xi_2\chi_2+\xi_3\chi_3\label{e1.5}
\end{equation}
where $e$, the identity element of $G$, is simply the origin
$(0,0,0)$. 

A representation $\T$ of $G$ in $L^2({\mathbb{R}},\mu)$, where $\mu$
is the Lebesgue measure, can be obtained by setting
\begin{equation}
\left(\T((\xi_1,\xi_2,\xi_3))f\right)(x)
=e^{-i\xi_3}e^{-ix\xi_2}f(x+\xi_1)\label{e1.6}
\end{equation}
It is easily seen that this is a continuous unitary representation of
$G$. 

The representation of $\G$, given by the
differential $d\T$ (with respect to the $L^2$-topology), is spanned by
the operators
\begin{equation}
T(\chi_1)\equiv X_1=\frac{d}{dx};\quad T(\chi_2)\equiv X_2=-ix;\quad
T(\chi_3)=X_3=iI\label{e1.7}
\end{equation}

The task at hand is to construct a rigged Hilbert space so that a
differentiable representation of $G$ maybe induced in the space $\Phi$
from the continuous unitary representation \eqref{e1.6} in $L^2$. To
that end, as a common invariant domain for the operator Lie algebra
\eqref{e1.7} we choose the Schwartz space $\S({\mathbb{R}})$, the
space of $C^\infty$-functions which decay at infinity faster than the
inverse of any polynomial. The definition \eqref{e1.6} shows that
$\S({\mathbb{R}})$  is invariant under the group representation
$\T$. We can now introduce the projective topology \eqref{2.5} on
$\S({\mathbb{R}})$ by means of the generators $X_1$, $X_2$, and $X_3$ of
\eqref{e1.7}:
\begin{equation}
 \l f\r_{n+1}^2=\l X_1f\r_n^2+\l X_2f\r_n^2+\l f\r_n^2,\qquad
 f\in{\S({\mathbb{R}})}\label{e1.8}
\end{equation}
This topology on $\S({\mathbb{R}})$ is equivalent to the more
customary one defined by the norms $\l
f\r_{m,n}=\sup_{x\in{\mathbb{R}}}|(\frac{d^n}{dx^n}x^mf)(x)|$. Thus,
$\S({\mathbb{R}})$ is complete under the topology \eqref{e1.8} and,
in fact, it is the projective limit of the scale of
Hilbert spaces $L^2({\mathbb{R}},\mu)\supset\H_1\supset\H_2\cdots$
where $\H_n$ is obtained by completing $\S({\mathbb{R}})$ with
respect to the norm $\l.\r_n$. Therefore, we have the RHS
\begin{equation}
\S({\mathbb{R}})\subset L^2({\mathbb
R},\mu)\subset\S({\mathbb{R}})^\times
\end{equation}
It is noteworthy that $\S({\mathbb{R}})$ is a nuclear space.

Proposition \ref{prop2.1} shows that the restriction of the
continuous unitary representation \eqref{e1.6} to the space
$\S({\mathbb{R}})$ yields therein a differentiable representation 
of the group \eqref{e1.1}. In fact, with respect to the norms
\eqref{e1.8},
\begin{equation}
\l\T(g)f\r_n\leq(1+|\xi_1|^2+|\xi_2|^2)^{n/2}\l f\r_n,\quad
n=0,1,2,\cdots,\ f\in\S({\mathbb{R}})\label{e1.9}
\end{equation}
where $g=(\xi_1,\xi_2,\xi_3)$. Further,
\begin{equation}
\lim_{g\rightarrow e}\l(\T(g)-I)f\r_n=0,\quad
n=0,1,2,\cdots,\ f\in\S({\mathbb{R}})\label{e1.10}
\end{equation}
and
\begin{eqnarray}
\lim_{\xi_1\rightarrow0}
\l\left(\frac{\T((\xi_1,0,0))-I}{\xi_1}-X_1\right)f\r_n&=&0   
\nonumber\\
\lim_{\xi_2\rightarrow0}
\l\left(\frac{\T((0,\xi_2,0))-I}{\xi_2}-X_2\right)f\r_n&=&0 
\nonumber\\
\lim_{\xi_3\rightarrow0}
\l\left(\frac{\T((0,0,\xi_3))-I}{\xi_3}-X_3\right)f\r_n&=&0,\quad
n=0,1,2,
\cdots,\ f\in\S({\mathbb{R}})\nonumber\\
\label{e1.11} 
\end{eqnarray}

As in the general case discussed in Proposition \ref{prop2.1}, the
proofs of \eqref{e1.9}--\eqref{e1.11} are by induction. The explicit
form of the factor $(1+|\xi_1|^2+|\xi_2|^2)^{n/2}$ in \eqref{e1.9}
follows from that of the functions  $f_{ij}$ of \eqref{2.15}, i.e.,
from $\T(g)X_1\T(g^{-1})=X_1+i\xi_2$, $\T(g)X_2\T(g^{-1})=X_2+i\xi_1$,   
or,
$f_{ij}(g)=\delta_{1i}\delta_{1j}+\delta_{2i}\delta_{2j}+
\delta_{3i}\delta_{3j}+\xi_2\delta_{1i}\delta_{3j}+
\xi_1\delta_{2i}\delta_{3j}$. 
In fact, the $f$'s are realized by the
(non-isomorphic) representation
$(\xi_1,\xi_2,\xi_3)\rightarrow(0,\xi_1,\xi_2)$ of $G$.

It is easily seen from \eqref{e1.9} that the differentiable
representation $\T$ extends to a continuous representation $\T_n$ for
every $n$. The generators of the one parameter subgroups $\T_n(\xi_1)$
and $\T_n(\xi_2)$ are, respectively,  the extensions to $\H_n$,  
by closure, of $X_1$ and $X_2$. As before, let us denote these two one parameter
subgroups in $\H_n$ by $\T(\xi_1,{\bar{X}_{1,n}})$ and
$\T(\xi_2,{\bar{X}_{2,n}})$. The classical Hille-Yosida theory can then be
applied to recover these one parameter subgroups from (the resolvents
of) their generators. 

Consider the one parameter subgroup $\T(\xi_2,{\bar{X}_{2,n}})$. From
\eqref{e1.9},
\begin{equation}
\l\T(\xi_2,{\bar{X}_{2,n}})f\r_n\leq(1+|\xi_2|^2)^{n/2}\l f\r_n,\quad 
f\in\H_n\label{e1.12}
\end{equation}
It is of type $\omega_n=0$, i.e.,
\begin{equation}
\omega_n=\inf_{|\xi_2|}\frac{1}{|\xi_2|}
\ln\l\T(\xi_2,{\bar{X}_{2,n}})\r_n=0,\quad 
n=0,1,2,\cdots\label{e1.13}
\end{equation}
Then, the resolvent operator $R(\lambda,{\bar{X}_{2,n}} )$ can be obtained
as, for $f\in\H_n$, 
\begin{eqnarray}
R(\lambda,{\bar{X}_{2,n}})f
&=&\int_0^\infty d\xi_2 e^{-\lambda\xi_2}\T(\xi_2,{\bar{X}_{2,n}})f,
\qquad\Re(\lambda)>0\nonumber\\
R(\lambda,{\bar{X}_{2,n}})f
&=&-\int_{-\infty}^0 d\xi_2 e^{-\lambda\xi_2}\T(\xi_2, {\bar{X}_{2,n}})f,
\qquad\Re(\lambda)<0\label{e1.14}
\end{eqnarray}
Also, the $R(\lambda,{\bar{X}_{2,n}})$ satisfy the equicontinuity
condition
\begin{equation}
\l
\left(R(\lambda,{\bar{X}_{2,n}})\right)^p\r_n\leq(|\lambda|-n)^{-p}\label{e1.15}
\end{equation}
for all $\lambda$ with $|\Re(\lambda)|>n$. Therefore, according to the
Hille-Yosida theory, the continuous group $\T(\xi_2,{\bar{X}_{2,n}})$ can
be recovered from the resolvent $R(\lambda,{\bar{X}_{2,n}})$ by means of
the limiting process \eqref{2.1.8}:   
\begin{eqnarray}  
\T(\xi_2,{\bar{X}}_n)\phi&=&\lim_{\lambda\rightarrow\infty}e^{-\lambda
\xi_2}\sum_{j=0}^\infty\frac{(\lambda\xi_2)^j}{j!}\left(\lambda
R(\lambda,{\bar{X}}_n)\right)^j\phi\quad{\rm for}\ \xi_2>0\nonumber\\
\T(\xi_2,{\bar{X}}_n)\phi&=&\lim_{\lambda\rightarrow-\infty}e^{-\lambda
\xi_2}\sum_{j=0}^\infty\frac{(\lambda\xi_2)^j}{j!}\left(\lambda
R(\lambda,{\bar{X}}_n)\right)^j\phi\quad{\rm for}\ \xi_2<0\nonumber\\
\end{eqnarray}
The differentiable one parameter
subgroup $\T(\xi_2,X_2)$ in $\Phi$ can then be obtained as the
projective limit of the continuous groups  $\T(\xi_2,{\bar{X}_{2,n}})$ in
$\H_n$. 

It is interesting to ask if the differentiable subgroup $\T(\xi_2,X_2)$
can be recovered from the resolvent operator $R(\lambda, X_2)$ in
$\Phi$, i.e., without appealing to the Banach space theory applied to
$\H_n$. Notice first that since
\begin{equation}
\omega=\sup_n\omega_n=0\nonumber
\end{equation}
where the $\omega_n$ are as in \eqref{e1.13}, the resolvent operator
$R(\lambda,X_2)$ is defined everywhere on the complex plane, except on
the imaginary axis, and it is given by integrals of the kind
\eqref{e1.14}. The formal integrals
\begin{eqnarray}
\int_0^\infty
d\xi_2e^{-\lambda\xi_2}e^{-ix\xi_2}f(x)=\frac{1}{\lambda+ix}f(x),\quad
\Re(\lambda)>0\nonumber\\
-\int_{-\infty}^0
d\xi_2e^{-\lambda\xi_2}e^{-ix\xi_2}f(x)=\frac{1}{\lambda+ix}f(x),\quad
\Re(\lambda)<0\label{e1.16}
\end{eqnarray}
which must coincide with the vector valued ones (which exist by the
above Hille-Yosida argument) show, for $f\in\S({\mathbb{R}})$, 
\begin{eqnarray}
R(\lambda, X_2)f(x)&=&\frac{1}{\lambda+ix}f(x)=\int_0^\infty
d\xi_2e^{-\lambda\xi_2}\T(\xi_2,X_2)f(x),\quad\Re(\lambda)>0\nonumber\\[0.1cm] 
R(\lambda, X_2)f(x)&=&\frac{1}{\lambda+ix}f(x)=-\int_{-\infty}^0
d\xi_2e^{-\lambda\xi_2}\T(\xi_2,X_2)f(x),\quad\Re(\lambda)<0\nonumber\\
\label{e1.17}
\end{eqnarray}
where the integrals are defined as the limit of a Riemann sum with
respect to Fr\'echet topology \eqref{e1.8} of $\S({\mathbb{R}})$. The
Hille-Yosida theory then implies that the operator $R(\lambda,X_2)$ is
an everywhere defined continuous operator in
$\S({\mathbb{R}})$. Alternatively, we could directly show, by
induction, that the
linear operator defined by the first equality in \eqref{e1.17} is such
an operator: 
\begin{equation}
\l R(\lambda,X_2)f\r_n\leq\left(\Pi_{i=0}^nc_i\right)^{1/2}\l
f\r_n\label{e1.18} 
\end{equation}
where $c_i=1+\Pi_{j=0}^{i-1}c_j$, $i=1,2,\cdots,n,$ and
$c_0=\frac{1}{|\lambda|^2}$. 

The relation \eqref{e1.18} also shows that $R(\lambda,X_2)$ extends to an
everywhere defined continuous operator in $\H_n$. This extension is
really the resolvent operator $R(\lambda,{\bar{X}_{2,n}})$ of
${\bar{X}_{2,n}}$, the closure of $X_2$ in $\H_n$-topology. Further, a
direct computation shows
\begin{equation}
\l
\left(R(\lambda, {\bar{X}_{2,n}})\right)^{p}\r_n\leq(|\lambda|-n)^{-p}\label{e1.19}
\end{equation}
for all $|\lambda|>n$ and $p=1,2,3,\cdots$. This is exactly the
relation \eqref{e1.15}, obtained there by applying the Hille-Yosida
theory to the $C_0$-group $\T(\xi_2,{\bar{X}_{2,n}})$ in $\H_n$.

The $\inf\{|\lambda|\}$, for which \eqref{e1.19} holds, strictly 
increases along the scale
$L^2({\mathbb{R}},\mu)\supset\H_1\supset\H_2\cdots$. This means that
the upper bound \eqref{2.1.11} does not exist for the $C_0$-group
$\T(\xi_2,X_2)$ in $\S({\mathbb{R}})$. That is, there exist no
$\beta\in{\mathbb{R}}$ such that $e^{-\beta\xi_2}\T(\xi_2,X_2)$ is
equicontinuous in $\S({\mathbb{R}})$ for $\xi_2\in{\mathbb{R}}$. 
Therefore, although the resolvent operator $R(\lambda, X_2)$ exists
for all $\lambda$ with $\Re(\lambda)\not=0$, the $C_0$-group
$\T(\xi_2,X_2)$ cannot be recovered from it by means of a limit
process akin to \eqref{2.1.8} in the
$\S({\mathbb{R}})$-topology. However, this
recovery can be done for each $\T(\xi_2,{\bar{X}_{2,n}})$ in $\H_n$,
and the differentiable group $\T(\xi_2,X_2)$ in $\S({\mathbb{R}})$ can
be obtained as the projective limit of the $\T(\xi_2,{\bar{X}_{2,n}})$
thus recovered. 

\subsection{Differentiable Representations of Groups in
$\boldsymbol{\Phi^\times}$}\label{sec2.4} 

Let $\T$ be a 
representation of a finite dimensional Lie group $G$ in
the space $\Phi$ of a rigged Hilbert space
$\Phi\subset\H\subset\Phi^\times$. Then, a representation $\V$
of $G$
can be defined in $\Phi^\times$ by way of the identity
\begin{equation}
\<\T(g)\phi|F\>=\<\phi|\V(g^{-1})F\>,\quad g\in G;\ \phi\in\Phi;\
F\in\Phi^\times\label{2.3.1}
\end{equation}
In other words,
\begin{equation}
\V(g^{-1})=(\T(g))^\times\label{dual group}
\end{equation}
where the right hand side denotes the operator dual to $\T(g)$. 
It is easy to verify that $\V$ is a homomorphism on $G$. Furthermore, 
if $\T$ is a continuous representation, $\V$ will also be a continuous
representation with respect to the weak* topology $\tau^\times$ in
$\Phi^\times$, and if $\T$ is differentiable, $\V$ will also be
differentiable. To see this, consider a one parameter
subgroup $\{e^{tx}\}$ in $G$  and its representation $\T(t,X)$ in 
$\Phi$. As in \eqref{2.3.1}, let us denote by 
$\V(t)$ the one parameter subgroup dual to $\T(t,X)$. 
If $\T$ is a differentiable representation, then for all
$\phi\in\Phi$,
$\lim_{t\rightarrow0}\frac{\left(\T(t,X)-I\right)}{t}\phi=X\phi$, and
thus,
\begin{eqnarray}
\<X\phi|F\>&=&\<\lim_{t\rightarrow0}\frac{\T(t,X)-I}{t}\phi,\
F\>\nonumber\\[.1cm]
&=&\lim_{t\rightarrow0}\<\frac{\T(t,X)-I}{t}\phi,\
F\>\nonumber\\[.1cm] 
&=&\lim_{t\rightarrow0}\<\phi,\ \frac{\V(-t)-I}{t}F\>
\nonumber\\[.1cm] 
&=&-\lim_{t\rightarrow0}\<\phi,\ \frac{\V(t)-I}{t}F\>\label{2.3.2}
\end{eqnarray}
where the second equality follows from the continuity of $F$ as an
antilinear functional on $\Phi$. 

The last equality in \eqref{2.3.2} shows that the
$\lim_{t\rightarrow0}\frac{\V(t)-I}{t}F$ exists everywhere in
$\Phi^\times$ with respect to the weak* topology $\tau^\times$. That
is, the dual representation $\V$, defined by \eqref{2.3.1}, is
differentiable in $\Phi^\times$ when $\T$ is differentiable in
$\Phi$. Further, since the operator $X^\times$ dual to $X$ is defined
by $\<X\phi, F\>=\<\phi,X^\times F\>,\ \phi\in\Phi,\ F\in\Phi^\times$,
we see from \eqref{2.3.2} that the generator of $\V(t)$ is
$-X^\times$, and we may thus denote the one parameter subgroup by
$\V(t,-X^\times)$. It is evident that the $\Phi^\times$-differential of
$\V$, evaluated at the identity element of $G$, furnishes a
representation $V$ of the Lie algebra $\G$, given explicitly by
\begin{equation} 
V(x)=-(T(x))^\times\quad x\in\G\label{2.3.3}
\end{equation}
where the $^\times$ on the right hand side denotes the dual operator
to $T(x)$. It is trivial to verify that the mapping $\G\rightarrow
V(\G)$ preserves the commutation relations
$[x_i,x_j]=c_{ijk}x_k$ in $\G$. 

\subsection{Example}\label{eg2}
Proposition \ref{prop2.1} shows that in a suitably constructed
rigged Hilbert space $\Phi\subset\H\subset\Phi^\times$, the
restriction $\T_\Phi$ of a continuous Lie group representation $\T$
in $\H$ furnishes a differentiable representation of the group in
$\Phi$. As seen in the previous section, by duality, there also exists
a differentiable representation of the group in the dual space
$\Phi^\times$, given in particular by $(\T(G))^\times$. It is
interesting to ask if every differentiable Lie group representation in
$\Phi$  necessarily arises as the restriction of a continuous
representation of the group in the kernel Hilbert space $\H$, or
equivalently, if every differentiable representation in $\Phi$ extends
to a continuous representation in $\H$. 
In this section we will construct a variant of the example considered
in Section~\ref{eg1} that shows that a differentiable representation
in the space $\Phi$ of an RHS need not extend to a continuous
representation in  the Hilbert space $\H$. However, this still 
leaves the case for nuclear spaces unanswered 
because our $\Phi$ here is not a nuclear vector space.

Consider again the Lie algebra $\G$ spanned by the $\chi_1$, $\chi_2$,
and $\chi_3$ of \eqref{e1.2}. The corresponding Lie group $G$ is
generated by the exponentiation of $\G$ as in \eqref{e1.5}. We can
obtain a representation of $\G$ in the Hilbert space
$\ell_2({\mathbb{C}})$ of square summable complex sequences
$\phi=(\phi_1,\phi_2,\phi_3,\cdots)$ by
the direct sum of the operator algebra \eqref{e1.4b}:
\begin{equation}
X_1=\sum_{n=1}^\infty\oplus n\chi_1,\quad 
X_2=\sum_{n=1}^\infty\oplus n\chi_2,\quad   
X_3=\sum_{n=1}^\infty\oplus n^2\chi_3\label{e2.1}
\end{equation}
i.e., $X_1\phi=(\phi_2,0,0,2\phi_5,0,0,3\phi_8,0,\cdots)$, etc.

The operators \eqref{e2.1} are unbounded on $\ell_2({\mathbb{C}})$. As
a common invariant dense domain for the $X_i$, and therewith for the
whole operator Lie algebra, we choose the subspace of rapidly
decreasing sequences, $\S=\{\phi: \phi\in\ell_2({\mathbb{C}});\ 
\lim_{|m|\rightarrow\infty}m^n\phi_m=0\ {\rm for}\ n=0,1,2,\cdots\}.$

To obtain an RHS, we introduce on  $\S$ a locally
convex topology by means of the scalar products
\begin{equation}
(\phi,\psi)_{n+1}=\sum_{i=1}^3\left(X_i\phi,X_i\psi\right)_n+
(\phi,\psi)_n\nonumber
\end{equation}
where $\phi,\ \psi\in\S$ and
$(\phi,\psi)_0=(\phi,\psi)=\sum_{m=1}^\infty\phi_m{\bar{\psi}}_m$, the
inner product in $\ell_2({\mathbb{C}})$. The ensuing norms are
\begin{equation}
\l\phi\r_{n+1}^2=\sum_{i=1}^3\l X_i\phi\r_n^2+\l\phi\r_n^2\label{e2.2}
\end{equation}

However, from \eqref{e1.4} and the definition \eqref{e2.1} of the
$X_i$, we have
\begin{equation}
X_iX_j=\delta_{1i}\delta_{2j}X_3\label{e2.3}
\end{equation}
Thus, the set of norms \eqref{e2.2} consists of only two elements:
\begin{eqnarray}
\l\phi\r_0^2&=&\l\phi\r^2=\sum_{m=1}^\infty|\phi_m|^2\nonumber\\[0.1cm]
\l\phi\r_1^2&=&\sum_{i=1}^3\l X_i\phi\r^2+\l\phi\r^2\label{e2.4}
\end{eqnarray}
The Hilbert space $\H_1$ which results from the completion of $\S$
under the norm $\l.\r_1$, its dual $\H_1^\times$, and
$\ell_2({\mathbb{C}})$ form the RHS
\begin{equation}
\Phi\equiv\H_1\subset\ell_2({\mathbb{C}})
\subset\H_1^\times\equiv\Phi^\times\label{e2.5}
\end{equation}

As mentioned earlier $\Phi$, being an infinite dimensional Hilbert
space, is not nuclear. 

In much the same way as the Lie algebra of \eqref{e1.3} integrates in
${\mathbb{R}}^3$  to a representation of the group $G$ of \eqref{e1.1}, the
operator Lie algebra spanned by the \eqref{e2.1} integrates in $\Phi$
to a differentiable representation of $G$:
\begin{equation}
\T(\xi_1,\xi_2,\xi_3)=I+\xi_1X_1+\xi_2X_2+\xi_3X_3\label{e2.6}
\end{equation}

That \eqref{e2.6} is a homomorphism on $G$ follows easily from
\eqref{e2.3} and \eqref{e1.1}. The continuity of
$\T(\xi_1,\xi_2,\xi_3)$ as an operator in $\Phi$ for each
$(\xi_1,\xi_2,\xi_3)\in G$, as well as the 
differentiability of the mapping $G\rightarrow\T(G)$ in $L(\Phi)$, 
follows from the continuity of the operators $X_i$ and the defining
relations \eqref{e2.6}.

Since the $X_i$ are not continuous in $\ell_2({\mathbb{C}})$,
\eqref{e2.6} does not yield a continuous representation of $G$ in the
central Hilbert space $\ell_2({\mathbb{C}})$ of the triad
\eqref{e2.5}. That is, the differentiable representation \eqref{e2.6}
in $\Phi$ does not extend to a continuous representation in
$\ell_2({\mathbb{C}})$. In fact, the operator Lie algebra spanned by
the $\{X_i\}$ of \eqref{e2.1} cannot be the differential of any
continuous representation of $G$ in $\ell_2({\mathbb{C}})$, be it in
the form \eqref{e2.6} or not, because none of basis elements $X_i$ is
integrable in $\ell_2({\mathbb{C}})$. To see this, first notice that
on the common invariant domain $\S$ for the $X_i$,
\begin{equation}
\frac{1}{\lambda^2}(\lambda+X_i)(\lambda-X_i)
=\frac{1}{\lambda^2}(\lambda-X_i)(\lambda+X_i)
=I,\quad\lambda\not=0\label{e2.7}
\end{equation}
If the resolvent operator $R(\lambda,X_i)$ exists for some non-zero
complex number $\lambda$, it must coincide with
$\frac{1}{\lambda^2}(\lambda+X_i)$ on $\S$. And for $\lambda=0$, the
range of $(\lambda-X_i)$ is not dense in
$\ell_2({\mathbb{C}})$. Therefore, the resolvent set of any of the
$X_i$ is empty, and the Hille-Yosida theory renders each $X_i$
non-integrable in $\ell_2({\mathbb{C}})$ to a $C_0$-group.

\section{Integrability of Operator Lie algebras in RHS}\label{sec3}

The Example \ref{eg2} 
motivates us to consider representations of Lie groups in $\Phi$
independently of possible corresponding representations of the group 
in $\H$.

Therefore, let us suppose that $T$ is a representation of a
$d$-dimensional ($d<\infty$) Lie algebra $\G$ in a complex Hilbert
space $\H$ by linear operators defined over a common, invariant dense
domain $\D$. Unlike in Section~\ref{sec2}, here we do not assume at the
outset that $T$ is the differential $d\T$ of a continuous Lie group
representation $\T$ in $\H$.

If $\{x_i\}_{i=1}^d$ is some basis for $\G$, then $T(x_i)$ furnishes a
basis for $T(\G)$, which is a finite dimensional subspace of the
algebra of endomorphisms on $\D$. We shall adopt the notation
$X=T(x),\ x\in\G$. Then, as in \eqref{2.4} and \eqref{2.5}, we may use
the operator algebra spanned by $\{X_i\}_{i=1}^d$ to define a locally
convex topology on $\D$ leading to an RHS
$\Phi\subset\H\subset\Phi^\times$, 
where $\Phi$ is the completion of $\D$ under the new locally convex
topology. By construction, every $X$ in $T(\G)$ is continuous as an
operator on $\Phi$. Thus, the mapping $T$ furnishes a representation
of $\G$ by continuous operators on $\Phi$. We shall denote this
operator Lie algebra in $\Phi$ also by $T(\G)$, unless there is room
for confusion. The problem we investigate in this section is the
integrability of $T$:

\begin{definition}\label{def3.1}
Let $G$ be a connected and simply connected Lie group and $\G$, its
Lie algebra. Let $T$ be a representation of $\G$ by (not necessarily
continuous) linear operators on a complex, locally convex, complete
topological vector space $\Psi$. $T$ is said to be integrable if
there exists a representation $\T$ of $G$ such that its differential
$d\T$, evaluated at the identity, contains $T$.
\end{definition}

In other words, integrability of $T$ means that for every $x\in\G$,
the operator $X=T(x)$ coincides on its domain of definition with the
generator of the one parameter subgroup $\T(e^{tx}),\
t\in{\mathbb{R}}$. The representation $\T$ is generally taken to be
continuous (Definition \ref{def2.1}).  The well known classical
results~\cite{yosida,komura,hille} then affirm that the generator of
the one parameter group $\T(e^{tx})$ is a densely defined, closed
operator in $\Psi$. When these generators are continuous, as seen in
Section~\ref{sec2}, the group representation $\T$ is not simply
continuous but differentiable (Definition \ref{def2.2}).

The integrability in the sense of  Definition \ref{def3.1} can be
viewed as an operator valued version of E.~Cartan's classic theorem
that every abstract Lie algebra is in fact the infinitesimal Lie
algebra of a Lie group. Integrating  operator Lie algebras has been a
subject of continued interest \cite{nelson,fsss,ref1,ref2,ref3,ref4}.
Among the earlier works are that of Nelson \cite{nelson} and of Flato
et.~al.~\cite{fsss}, where  primarily the integration of operator Lie
algebras into unitary group representations in Hilbert spaces is
investigated. The problem is also studied for more general cases 
such as Banach spaces and other locally
convex spaces \cite{ref1,ref2,ref3,ref4}. Some of these developments make
use of a good deal of geometric notions, whereas \cite{nelson} and
\cite{fsss} mainly employ techniques of functional analysis. Since the
locally convex spaces in rigged Hilbert spaces have a particular
topological structure as the projective limit of a scale of Hilbert
spaces \eqref{2.4}, for the purposes of this paper what is mostly
relevant is the constructions in \cite{nelson} and, in particular,
\cite{fsss} for Hilbert spaces; our main technical result (Theorem
\ref{th3.1}) is an immediate extension of \cite{fsss}.   Therefore,
we shall not review here in detail the treatments of
\cite{ref1,ref2,ref3,ref4} which deal with various aspects of the
integrability problem in Banach and other locally convex spaces.

The centrally significant theoretical feature for the unitary
representations is the existence of a large class of analytic vectors
for the representation $T(\G)$. In particular, Nelson proved
\cite{nelson} that if the Laplacian $\Delta=-\sum_{i=1}^d X_i^2$ with
respect to some basis $\{X_i\}$ of $T(\G)$ is essentially self-adjoint
for a Lie algebra representation $T$ by skew symmetric operators
defined on a common invariant dense domain in a Hilbert space, then
$T$ is integrable to a unique unitary representation of $G$. A
generalization of Nelson's integrability criterion for unitary
representations was achieved by M.~Flato et al.~(FSSS). They
proved~\cite{fsss} that a Lie algebra isomorphism by skew symmetric
operators in a Hilbert space is integrable to a unique unitary
representation of $G$ if there exists an invariant common dense domain
of vectors analytic for some basis $\{X_i\}$ of the operator Lie
algebra. That is, these vectors are assumed to be analytic for each
$X_i$ separately, but not necessarily for the  whole Lie
algebra. Thus, the FSSS theory provides less stringent integrability
condition than Nelson's.

Furthermore, the FSSS theory has the interesting feature that it can
be naturally generalized to continuous group
representations in more general, complete locally convex
spaces~\cite{fsss}. This generalization is achieved, however,
contingent to the assumption, which supplements the ones on the
existence of analytic vectors, that the closure of each basis element
$\bar{X_i}$  generates a one parameter subgroup. Although this
requirement is redundant for skew symmetric operators in Hilbert
spaces, the integrability problem for an operator in a general locally
convex space into a continuous one parameter group is considerably
more complex, especially when the group is not globally equicontinuous
in the parameter. Such was the case considered in 
Example \ref{eg1}. 

In this section, we propose an adaptation of the FSSS theory for Lie
group representations in rigged Hilbert
spaces.  As mentioned above, for our purposes, the FSSS theory
provides the most convenient and immediate starting point.
Suppose then an RHS $\Phi\subset\H\subset\Phi^\times$ has been
built so as to yield an isomorphism $T$ of a Lie algebra by continuous
linear operators in $\Phi$. Thus, the integrability of $T$ amounts to
finding a true anti derivative for $T$, i.e., a group representation
$\T$ such that $d\T=T$ everywhere in $\Phi$. That is, the group
representation $\T$ is differentiable, not just continuous as
considered in~\cite{fsss}. Our main technical result is
that the differentiability of $\T$ allows us to remove the assumption
on the existence of analytic vectors in the FSSS theory. This absence
of the need for analytic vectors may make matters considerably simpler
in applications.

\begin{theorem}\label{th3.1}
Let $\Phi\subset\H\subset\Phi^\times$ be a rigged Hilbert space and
$L(\Phi)$, the space of continuous linear operators in $\Phi$ equipped
with the strong operator topology. Let $\G$ be a Lie algebra of dimension
$d<\infty$ and $G$, the connected and simply connected Lie group with
$\G$ as its Lie algebra. Suppose $T:\ \G\rightarrow T(\G)\subset
L(\Phi)$ is an isomorphism on $\G$, and suppose that there exists a
basis $\{x_i\}_{i=1}^d$ for $\G$ such that each $X_i\equiv T(x_i),\ 
i=1,2,3,\cdots,d,$ generates a one parameter group in $\Phi$. Then $T$ 
is integrable to a unique differentiable representation of $G$.
\end{theorem}

Before we present the proof of Theorem~\ref{th3.1}, we shall consider
some preliminary facts and identities from Lie group theory and
formulate their operator valued analogues in $L(\Phi)$.

\subsection{Lie Algebra Preliminaries}\label{sec3.2}

Let $G$, $\G$, and $\{x_i\}$ be as defined in
Theorem~\ref{th3.1}. Then, a convex neighborhood $W$ of the identity
$e$ of $G$ can be chosen such that any $g\in W$ can be written as
\begin{equation}
g=e^{t_1(g)x_1}e^{t_2(g)x_2}\cdots e^{t_d(g)x_d}\label{3.2.1}
\end{equation}
The coordinate functions of the second kind
\begin{equation}
g\rightarrow(t_1(g),t_2(g),\cdots,t_d(g))\label{3.2.2} 
\end{equation}
furnish a local chart over $W$. Since $W$ is chosen to be convex, we
have $e^{tx}e^y\in W$ whenever $e^y\in W$, $e^xe^y\in W$, and $0\leq
t\leq1$. Thus, for any $e^x\in W$ and $0\leq t\leq1$,
\begin{equation}
e^{tx}=e^{t_1(t)x_1}e^{t_2(t)x_2}\cdots e^{t_d(t)x_d}\label{3.2.3}
\end{equation}
where we use the simpler notation $t_i(e^{tx})\rightarrow t_i(t)$. 

By applying the chain rule of differentiation on \eqref{3.2.3}, we obtain 
the Lie algebra identities
\begin{eqnarray}
x&=&\frac{dt_1}{dt}x_1+\cdots+\frac{dt_d}{dt}
\Int(t_1x_1)\cdots{\rm Int}(t_{d-1}x_{d-1})x_d\nonumber\\[0.0703cm]
x&=&{\rm Int}(-t_dx_d)\cdots{\rm
Int}(-t_2x_2)x_1\frac{dt_1}{dt}+\cdots+x_d\frac{dt_d}{dt}\label{3.2.8}
\end{eqnarray}
where $\Int(tx)y=e^{tx}ye^{-tx}$, the inner automorphism on $\G$
induced by the elements of $G$. We shall also make use of the well-known
formula
\begin{equation}
 e^{tx}ye^{-tx}=\Int(tx)y
=\sum_{n=0}^\infty\frac{1}{n!}\left(\ad({tx})\right)^ny\label{3.2.7}
\end{equation}
where $(\ad(x))^ny=[x,(\ad(x))^{n-1}y]$ and $(\ad(x))^0y=y$. The
series on the right 
hand side of \eqref{3.2.7} converges in the usual Euclidean topology of
$\G$.

Finally, for some $x\in\G$ and $g\in W$ such that $e^xg\in W$, we have
by the convexity of $W$, $e^{tx}g\in W$ for $0\leq t\leq 1$. Thus, by
\eqref{3.2.3},
\begin{equation}
e^{tx}g=e^{\a_1(t)x_1}e^{\a_2(t)x_2}\cdots e^{\a_d(t)x_d}\label{3.2.8a}
\end{equation}
where the $\a_i$ are analytic in $t$ as they are simply given by the
coordinate functions $t_i$ of \eqref{3.2.3} as $\a_i(t)=t_i(e^{tx}g).$ 
This yields the identities~\cite{fsss},
\begin{eqnarray}
x&=&\frac{d\a_1}{dt}x_1+\cdots+
\frac{d\a_d}{dt}\Int(\a_1x_1)\cdots\Int(\a_{d-1}x_{d-1})x_d
\nonumber\\[0.0703cm]  
g^{-1}xg&=&\Int(-\a_dx_d)\ldots\Int(-\a_2x_2)x_1\frac{d\a_1}{dt}
+\cdots+x_d\frac{d\a_d}{dt}\label{3.2.9}
\end{eqnarray}

\subsection{$\boldsymbol{L(\Phi)}$ Analogues}\label{sec3.2.3}

Let us denote the image of $\Int(x)y$ under the isomorphism $T$ by
$\Int(X)Y$, i.e., $\Int(X)Y\equiv T(\Int(x)y)$.

\begin{proposition}\label{prop3.1}
\begin{equation}
(\Int(tX)Y)\phi
=T(e^{tx}ye^{-tx})\phi
=\sum_{n=0}^\infty\frac{1}{n!}(\ad(tX))^nY\phi,\quad\phi\in\Phi
\label{3.2.10}
\end{equation}
where $(\ad(X))^nY\phi=X(\ad(X))^{n-1}Y\phi-(\ad(X))^{n-1}YX\phi$. 
\end{proposition}
\noindent {\footnotesize PROOF:}\\
The first equality follows trivially from the above definition and the
Lie algebra identity \eqref{3.2.7}. What needs to be shown is the
convergence of the series in $L(\Phi)$ and that its limit is
$\Int(tX)Y$. But this is trivial from \eqref{3.2.7} and the continuity
of the mapping $T:\ \G\rightarrow L(\Phi)$ in the strong operator
topology of $L(\Phi)$.\hfill$\Box$

\vskip .5cm 
Recall that the basis $\{x_i\}$ is chosen in $\G$ so that each
$X_i=T(x_i)$ integrates to a one parameter group of
operators in $\Phi$. Let $\T(t,X_i)$ be this group. Then, by the
continuity of $X_i$,
\begin{equation}
\frac{d}{dt}\T(t,X_i)\phi=X_i\T(t,X_i)\phi
=\T(t,X_i)X_i\phi,\quad\phi\in\Phi\label{3.2.11}
\end{equation}
Further, since $\Phi$ is a Fr\'echet space, $\T(t,X_i)$ is 
locally equicontinuous, i.e., 
for any compact interval
$I\subset\mathbb{R}$ and any $n$, there exists 
some $m$ such that
\begin{equation}
\l\T(t,X_i)\phi\r_n\leq\l\phi\r_m,\qquad t\in I,\ \phi\in\Phi\label{3.2.12}
\end{equation} 
The relations \eqref{3.2.11} and \eqref{3.2.12}  
are among the standard results of the theory of
one-parameter groups in locally convex spaces~\cite{komura}.

\begin{proposition}\label{prop3.2}
\begin{equation}
\frac{d}{dt}\T(t,X_i)\T(t,X_j)\phi=
\T(t,X_i)(X_i+X_j)\T(t,X_j)\phi,\quad\phi\in\Phi\label{3.2.13}
\end{equation}
\end{proposition}

\noindent {\footnotesize PROOF:}\\
{\scriptsize 
\begin{eqnarray}
\lefteqn{
\l\frac{d}{dt}\T(t,X_i)\T(t,X_j)\phi-\T(t,X_i)(X_i+X_j)\T(t,X_j)\phi\r_n}
\nonumber\\[0.0703cm]
&&=\lim_{s\rightarrow0}\l\frac{\T(t+s,X_i)\T(t+s,X_j)
-\T(t,X_i)\T(t,X_j)}{s}\phi
-\T(t,X_i)(X_i+X_j)\T(t,X_j)\phi\r_n\qquad\qquad\qquad\qquad
\nonumber\\[0.0703cm]
&&\leq\lim_{s\rightarrow0}\l\T(t+s,X_i)\frac{\T(t+s,X_j)-\T(t,X_j)}{s}
-\T(t+s,X_i)X_j\T(t,X_j)\phi\r_n\nonumber\\
&&\quad+\lim_{s\rightarrow0}\l\T(t+s,X_i)X_j\T(t,X_j)\phi
-\T(t,X_i)X_j\T(t,X_j)\phi\r_n\qquad\qquad\nonumber\\[0.0703cm]
&&\quad\quad+\lim_{s\rightarrow0}\l\frac{\T(t+s,X_i)-\T(t,X_i)}{s}\T(t,X_j)\phi
-\T(t,X_i)X_i\T(t,X_j)\phi\r_n\nonumber\\[0.0703cm]
&&\leq\lim_{s\rightarrow0}\l\frac{\T(t+s,X_j)-\T(t,X_j)}{s}-X_j\T(t,X_j)\phi\r_m\nonumber\\[0.0703cm]
&&\quad+\lim_{s\rightarrow0}\l\T(t+s,X_i)X_j\T(t,X_j)\phi
-\T(t,X_i)X_j\T(t,X_j)\phi\r_n\nonumber\\[0.0703cm]
&&\quad\quad+\lim_{s\rightarrow0}\l\frac{\T(t+s,X_i)-\T(t,X_i)}{s}\T(t,X_j)\phi
-\T(t,X_i)X_i\T(t,X_j)\phi\r_n\nonumber
\end{eqnarray}}
The first term in the last inequality follows from the local
equicontinuity of $\T$, \eqref{3.2.12}. Since each term on right
hand side tends to zero, we have \eqref{3.2.13}. Notice that we needed only
the local equicontinuity of $\T(t,X_i)$ but not that of $\T(t,X_j)$
for \eqref{3.2.13} to hold.\hfill$\Box$
\vskip .5cm
Relations \eqref{3.2.10} and \eqref{3.2.13} can be combined to obtain an
$L(\Phi)$ analogue of the Lie algebra identity \eqref{3.2.7}:

\begin{proposition}\label{prop3.3}
For any two basis elements $X_i$ and $X_j$ of $T(\G)$, the equality
\begin{equation}
\T(t,X_i)X_j\T(-t,X_i)\phi
=\sum_{n=0}^\infty\frac{1}{n!}\left(\ad(tX_i)\right)^nX_j\phi\label{3.2.14}
\end{equation}
holds for all $\phi\in\Phi$. The series here is defined as in
\eqref{3.2.10} in the strong operator topology.
\end{proposition}
\noindent{\footnotesize PROOF:}\\
The proposition is clearly true for $t=0$. Next, by the continuity of
the linear mapping
$T:\ \G\rightarrow L(\Phi)$, we have
\begin{equation}
\frac{d}{dt}T(e^{tx_i}x_je^{-tx_i})\phi
=T\left(\frac{d}{dt}(e^{tx_i}x_je^{-tx_i})\right)\phi
=T(e^{tx_i}(\ad({x_i})x_j)e^{-tx_i})\phi
\label{3.2.15}
\end{equation}
But, by \eqref{3.2.7}, 
\begin{equation}
e^{tx_i}(\ad{x_i})x_je^{-tx_i}
=(\ad{x_i})(e^{tx_i}x_je^{-tx_i})
=(\ad{x_i})\sum_{n=0}^\infty\frac{1}{n!}(\ad({tx_i}))^nx_j
\nonumber
\end{equation}
Thus, again by the continuity of $T:\ \G\rightarrow L(\Phi)$,
\begin{equation} 
T\left((\ad({x_i}))(e^{tx_i}x_je^{-tx_i})\right)\phi
=(\ad{X_i})\sum_{n=0}^\infty\frac{1}{n!}(\ad({tX_i}))^nX_j\phi
\label{3.2.16}
\end{equation}

Therefore, from \eqref{3.2.10}, \eqref{3.2.15}, and \eqref{3.2.16}, we have
\begin{equation}
\frac{d}{dt}\sum_{n=0}^\infty
\frac{1}{n!}\left(\ad(tX_i)\right)^nX_j\phi
=(\ad{X_i})\sum_{n=0}^\infty\frac{1}{n!}(\ad({tX_i}))^nX_j\phi,
\quad\phi\in\Phi\label{3.2.17}
\end{equation}
Now, from \eqref{3.2.13},
\begin{eqnarray}
\frac{d}{dt}\
\T(t,X_i)X_j\T(-t,X_i)\phi
&=&\T(t,X_i)\left(\ad(X_i)X_j\right)\T(-t,X_i)\phi\nonumber\\
&=&(\ad({X_i}))\left(\T(t,X_i)X_j\T(-t,X_i)\right)\phi,
\quad\phi\in\Phi\nonumber\\
\label{3.2.18}
\end{eqnarray}

Equalities \eqref{3.2.17} and \eqref{3.2.18} yield the $L(\Phi)$ valued
differential equation
\begin{equation}
\frac{d}{dt}\ u(t)\phi=(\ad(X_i))u(t)\phi\quad\phi\in\Phi\label{3.2.19}
\end{equation}
where
\begin{equation}
u(t)=\T(t,X_i)X_j\T(t,X_i)-\displaystyle{\sum_{n=0}^\infty}\frac{1}{n!}
\left(\ad(tX_i)\right)^nX_j\label{3.2.20}
\end{equation}
Thus $u(0)=0$. We can employ a technique used in~\cite{fsss}, Equation
(12), redefined here with respect to the $L(\Phi)$ topology, to show
that the solution $u(t)$ to \eqref{3.2.19} is identically equal to
zero. To that end, consider the function
$v(s)\phi=\T(t-s,X_i)u(s)\T(-t+s,X_i)\phi$, where $u(s)$ is as in
\eqref{3.2.20}.

From \eqref{3.2.10} and \eqref{3.2.12}, $u(s)$ is locally equicontinuous
in $s$. Hence, by Proposition~\ref{prop3.2},
\begin{equation}   
\frac{dv(s)}{ds}\phi
=-X_iv(s)\phi+(\ad{X_i})v(s)\phi+v(s)X_i\phi
=0\label{3.2.21}
\end{equation}
i.e., $v(s)$ is independent of $s$. Therefore,
\begin{equation}
u(t)=v(t)=v(0)=\T(t,X_i)u(0)\T(-t,X_i)=0\nonumber
\end{equation}
and \eqref{3.2.20} gives \eqref{3.2.14}\hfill$\Box$

In summary,
\begin{equation}
\T(t,X_i)X_j\T(-t,X_i)\phi=(\Int(tX_i)X_j)\phi
=\sum_{n=0}^\infty\frac{t^n}{n!}(\ad({X_i}))^nX_j\phi\label{3.2.22}
\end{equation}
\noindent{\bf Remark:}
This equality is similar to Equation (8) of~\cite{fsss}. However, our
assumptions as well as proof technique are different.
\vskip .5cm
Further, from \eqref{3.2.8} and \eqref{3.2.9}, 
we also have the $L(\Phi)$ valued the Lie algebra
identities: 
\begin{eqnarray}
X&=&\frac{dt_1}{dt}X_1+\cdots+\frac{dt_d}{dt}
\Int(t_1X_1)\cdots{\rm Int}(t_{d-1}X_{d-1})X_d\nonumber\\[0.0703cm]
X&=&{\rm Int}(-t_dX_d)\cdots{\rm
Int}(-t_2X_2)X_1\frac{dt_1}{dt}+\cdots+X_d\frac{dt_d}{dt}\nonumber\\[0.0703cm]
X&=&\frac{d\a_1}{dt}X_1+\cdots+\frac{d\a_d}{dt}
\Int(\a_1X_1)\cdots{\rm Int}(\a_{d-1}X_{d-1})X_d\label{3.2.23}
\end{eqnarray}

All the technical preliminaries are now in place for the proof of
Theorem~\ref{th3.1}

\subsection{Proof of Theorem \ref{th3.1}}
Let $W$ be as defined Section~\ref{sec3.2}. Then, for 
any $g\in W$ (i.e., of the form \eqref{3.2.1}) we define an $L(\Phi)$
element 
$\T(g)$ by 
\begin{equation}
\T(g)=\T(t_1(g),X_1)\T(t_2(g),X_2)\ldots\T(t_d(g),X_d)\label{3.2.24}
\end{equation}
Being the composition of finitely many continuous linear operators, 
$\T(g)$ is a continuous linear operator. Next, if $x\in\G$ is such
that $e^{x}g\in W$, then for $0\leq t\leq1$  by way of \eqref{3.2.3} and
\eqref{3.2.8a},
\begin{eqnarray}
\T(e^{tx})&=&\T(t_1(t),X_1)\T(t_2(t),X_2)\cdots\T(t_d(t),X_d)
\nonumber\\[0.0703cm] 
\T(e^{tx}g)&=&\T(\a_1(t),X_1)\T(\a_2(t),X_2)\cdots\T(\a_d(t),X_d)\label{3.2.25}
\end{eqnarray}

Since each $\T(t,X_i)$ is locally equicontinuous, repeated
applications of Proposition~\ref{prop3.2} and 
Proposition~\ref{prop3.3} (\eqref{3.2.22} in
particular) on the first equality in
\eqref{3.2.25} yield, for all $\phi\in\Phi$,
{\footnotesize 
\begin{eqnarray} 
\frac{d}{dt}\T(e^{tx})\phi&=&\left(\frac{dt_1}{dt}X_1
+\cdots+\frac{dt_d}{dt}
\Int(t_1X_1)\cdots\Int(t_{d-1}X_{d-1})X_d\right)\T(e^{tx})\phi
\nonumber\\[0.0703cm]
\frac{d}{dt}\T(e^{tx})\phi&=&\T(e^{tx})\left(\Int(-t_dX_d)
\cdots\Int(-t_2X_2)X_1\frac{dt_1}{dt}
+\cdots+X_d\frac{dt_d}{dt}\right)\phi\nonumber
\label{3.2.26}
\end{eqnarray}}
Thus, with \eqref{3.2.23}, we have, for all $\phi\in\Phi$,
\begin{equation}
\frac{d}{dt}\ \T(e^{tx})\phi=X\T(e^{tx})\phi= \T(e^{tx})X\phi,\qquad
\phi\in\Phi\label{3.2.27} 
\end{equation}

This shows the differentiability of $\T(e^{tx})$ in the neighborhood
$W$ of the identity of $G$. 

The same application on the second equality in \eqref{3.2.25}, together with
\eqref{3.2.23}, gives
\begin{equation}
\frac{d}{dt}\T(e^{tx}g)\phi
=X\T(e^{tx}g)\phi\quad\phi\in\Phi\label{3.2.28}
\end{equation} 

Next, for $0\leq s\leq t\leq1$, the vector valued function
\begin{equation}
f(s)\phi=\T(e^{sx})\T(e^{(t-s)x}g)\phi\nonumber
\end{equation}
can be differentiated, as in \eqref{3.2.13}, because $\T(s^{sx})$ and
$\T(e^{(t-s)x}g)$ are both locally equicontinuous in $s$. Thus,
\begin{equation}
\frac{d}{ds}\
f(s)\phi=X\T(e^{sx})\T(e^{(t-s)x}g)\phi
-\T(e^{sx})X\T(e^{(t-s)x}g)\phi=0\label{3.2.29}
\end{equation}
That is, $f(s)\phi$ is independent of $s$, and so,
\begin{equation}
f(0)\phi=\T(e^{tx}g)\phi=f(t)\phi
=\T(e^{tx})\T(g)\phi\label{3.2.30}
\end{equation}
This shows that the mapping $W\rightarrow\T(g)$ defined by
\eqref{3.2.24} is a homomorphism on $W$. 

Recall that $G$ was assumed to be the connected and simply connected
Lie group with $\G$ as its Lie algebra. Thus, an arbitrary element $g$
of $G$ can be written as a product of finitely many elements of
$W$. Consequently, the homomorphism $\T:\ W\rightarrow L(\Phi)$ given
by \eqref{3.2.24} can be extended from $W$ to the entire group manifold,
and the simply connectedness of $G$ assures that this extension 
is well defined for all $g\in G$. From \eqref{3.2.27} and the
analyticity of the multiplication in $G$, it follows that the above
extension yields a differentiable representation of $G$ in $\Phi$. It
is straightforward to verify, by way of \eqref{3.2.28}, that the
differential $d\T|_e$ coincides with the Lie algebra representation
given at the outset, $T$.\\
This concludes the proof of Theorem~\ref{th3.1}\hfill$\Box$

As an immediate consequence of the theorem, we have the following
corollary:
\begin{corollary}\label{cor3.1}    
Under the assumptions of Theorem \ref{th3.1}, the dual Lie algebra
representation in $\Phi^\times$, 
defined by $T^\times(x)=-\left(T(x)\right)^\times,\ x\in\G$, is
integrable.
\end{corollary}
\noindent{\footnotesize  PROOF:}\\
If $T$ is integrable to the differentiable representation $\T$ in
$\Phi$, then as defined by \eqref{dual group}, there exists a 
differentiable representation $\V$ in $\Phi^\times$. The weak*
differential $d\V$ of $\V$ is precisely $-(T(\G))^\times$.\hfill$\Box$ 
\vskip .5cm

Example \ref{eg2} led us to the conclusion that not every
differentiable Lie group representation in $\Phi$ comes about as the
restriction of a continuous representation of the group in $\H$. The
following proposition allows us to determine if such is the case for a
given differentiable representation $\Phi$ of an RHS.

\begin{proposition}\label{prop3.4}
Let $\G$ and $G$ be as in Theorem~\ref{th3.1}, and let $\T$ be a
differentiable representation of $G$ in the space $\Phi$ of a rigged
Hilbert space $\Phi\subset\H\subset\Phi^\times$. Suppose there exists
a basis $\{X_i\}_{i=1}^d$ for the operator Lie algebra $T(\G)$ such
that each one parameter subgroup $\T(t,X_i)$ extends to a continuous
one parameter subgroup in $\H$. Then the differentiable representation
$\T$ extends to a continuous representation of $G$ in $\H$.
\end{proposition}
\noindent{\footnotesize PROOF:}\\
Since the extension of the one parameter subgroup $\T(t,X_i)$ in
$\H$ is generated by the $\H$-closure $\bar{X_i}$ of the generator
$X_i$, let us denote it by $\T(t,\bar{X_i})$. Now, for $g\in W$,
where $W$ is as in the proof of Theorem~\ref{th3.1}, define 
\begin{equation}
\T_{\H}(g)\phi=\T(t_1(g),\bar{X_i})\cdots\T(t_d(g),\bar{X_d})
\phi\quad\phi\in\H,\ g\in W\label{3.2.31}
\end{equation}
It is clear that $\T_{\H}(g)$ is a continuous linear operator in $\H$
for each $g\in W$. Since $\T_{\H}(g)$ coincides with $\T(g)$ of
\eqref{3.2.24} on $\Phi$ and since $\Phi$ is dense in $\H$, the mapping
$\T_{\H}:\ W\rightarrow\B(\H)$ of \eqref{3.2.31} is a homomorphism on $W$. 
For $x\in\G$ such that $e^{tx}\in W,\ 0\leq t\leq1$,  we have 
\begin{equation}
\T_{\H}(e^{tx})\phi
=\T(t_1(t),\bar{X_1})\T(t_2(t),\bar{X_2})\cdots
\T(t_d(t),\bar{X_d})\phi\quad \phi\in\H\label{3.2.32}
\end{equation}
which shows that $\T_{\H}:\ W\rightarrow\B(\H)$ is continuous on
$W$. As in the proof of Theorem~\ref{th3.1}, the connectedness and
simply connectedness of $G$ permits a well defined extension of
$\T_{\H}$ from $W$ to the entire $G$ to yield a continuous
representation of $G$ in $\H$. \hfill$\Box$

In view of Proposition~\ref{prop2.1}, $\T$ in $\Phi$ is then the 
projective limit of continuous representations of $G$ in a scale of
Hilbert spaces $\H\supset\H_1\supset\H_2\cdots$.  

\section{Concluding Remarks}\label{sec4}
This paper studies some aspects of differentiable representations of
finite dimensional Lie groups in rigged Hilbert spaces. In particular,
it is shown (Proposition~\ref{prop2.1}) that, for a suitably
constructed rigged Hilbert space, such a representation can always be
obtained from a continuous representation of the group defined in a
Hilbert space. Further, conditions are specified (Theorem~\ref{th3.1})
under which a given Lie algebra representation in a Hilbert space may
be integrated to an RHS representation of the corresponding Lie group.
It is worthwhile to point out that, in a suitable RHS
$\Phi\subset\H\subset\Phi^\times$, such integrability may be possible
in $\Phi$ even when the given Hilbert space representation of the Lie
algebra is not integrable in the Hilbert space $\H$ itself
(Proposition \ref{prop3.4}). 

Lie groups and Lie algebras play an essential role in many quantum
mechanical theories. Building a part of the theoretical framework for
handling Lie group and algebra representations in the RHS formulation
of quantum mechanics is the primary goal of this paper. In addition,
as pointed out in the Introduction, recent applications of the
formalism to characterize relativistic resonances and unstable
particles involve intricacies of the representations of Lie groups (and
subsemigroups thereof) in RHS. In the developments achieved in \cite{rgv}, 
the space $\Phi$, and therewith the RHS
$\Phi\subset\H\subset\Phi^\times$, is built so that a 
differentiable representation of the Poincar\'e semigroup (introduced
in Section \ref{sec1}) can be 
obtained in $\Phi$ from a unitary representation of the Poincar\'e group in
$\H$. In particular, these constructions employ 
Proposition \ref{prop2.1} to obtain a differentiable representation 
of the homogeneous Lorentz group in $\Phi$. Further, the construction
of $\Phi$ is achieved so that the momentum operators $P_\mu$ do not
generate one parameter groups in $\Phi$, and thus (Theorem
\ref{th3.1}) the differentiable representation of the Poincar\'e
semigroup in $\Phi$ does not extend to a representation of the entire 
Poincar\'e group. Motivation for the mathematical developments
presented in this paper partly comes from the theory of relativistic
resonances and unstable particles proposed in \cite{rgv}.  

\section*{Acknowledgements}
The authors are  grateful to the Erwin Schr\"odinger Institute for
Mathematical Physics, Vienna, where they were guests while a part of
this paper was written. They are also grateful for the financial support
from the Welch Foundation.

\end{document}